\documentclass[12pt,aps,ams,amsfonts,nofootinbib]{revtex4}

\usepackage[dvips]{graphics}

\usepackage{graphicx}
\usepackage{amssymb}
\usepackage{subfigure}
\usepackage{stackrel}
\usepackage{enumitem}
\usepackage{amsmath}
\usepackage{amsfonts}
\usepackage{bm}
\usepackage{color}
\usepackage{verbatim} 
\usepackage{tcolorbox}
\usepackage[normalem]{ulem}
\usepackage{hyperref}
\usepackage{comment}

\newlist{subquestion}{enumerate}{1}
\setlist[subquestion,1]{label=(\alph*)}
\newcommand{\vect}[1]{\ensuremath{\bm{{#1}}}}
\newcommand{\ket}[1]{\ensuremath{\left|{#1}\right\rangle}}
\newcommand{\bra}[1]{\ensuremath{\left\langle{#1}\right |}}
\newcommand{\operdef}[2]{\ensuremath{|{#2}\rangle \langle{#1}|}}

\newcommand{\beq}{\begin{equation}}
\newcommand{\eeq}{\end{equation}}
\newcommand{\bse}{\begin{subequations}}
	\newcommand{\ese}{\end{subequations}}\newcommand{\bea}{\begin{eqnarray}}
\newcommand{\eea}{\end{eqnarray}}
\newcommand{\bit}{\begin{itemize}}
	\newcommand{\eit}{\end{itemize}}
\newcommand{\bpmatrix}{\begin{pmatrix}}
	\newcommand{\epmatrix}{\end{pmatrix}}

\newcommand{\be}{\begin{equation}}
\newcommand{\ee}{\end{equation}}
\newcommand{\ben}{\begin{eqnarray}}
\newcommand{\een}{\end{eqnarray}}

\begin{document}

\title{Non-commutative measure of quantum correlations under local operations}

\author{D. G. Bussandri$^{1,2}$ , A. P. Majtey$^{1,3}$, A. Valdés-Hernández$^4$ }
\affiliation{$^1$Facultad de Matem\'atica, Astronom\'{\i}a, F\'{\i}sica y Computaci\'on, Universidad Nacional de C\'ordoba, Av. Medina Allende s/n, Ciudad Universitaria, X5000HUA C\'ordoba, Argentina}
\affiliation{$^2$Consejo Nacional de Investigaciones Cient\'{i}ficas y T\'ecnicas de la Rep\'ublica Argentina, Av. Rivadavia 1917, C1033AAJ, CABA, Argentina}
\affiliation{$^3$ Instituto de F\'isica Enrique Gaviola, Consejo Nacional de Investigaciones Cient\'{i}ficas y T\'ecnicas de la Rep\'ublica Argentina, Av. Medina Allende s/n, X5000HUA, C\'{o}rdoba, Argentina}
\affiliation{$^4$ Instituto de F\'{\i}sica, Universidad Nacional Aut\'{o}noma de M\'{e}xico, Apartado Postal 20-364, Ciudad de M\'{e}xico, Mexico.}


\begin{abstract}
We study some desirable properties of recently introduced measures of quantum correlations based on the amount of non-commutativity quantified by the Hilbert-Schmidt norm (Sci Rep 6:25241, 2016, and Quantum Inf. Process. 16:226, 2017). Specifically, we show that: 1) for any bipartite ($A+B$) state, the measures of quantum correlations with respect to subsystem $A$ are non-increasing under any Local Commutative Preserving Operation on subsystem $A$, and 2) for Bell diagonal states, the measures are non-increasing under arbitrary local operations on $B$. Our results accentuate the potentialities of such measures, and exhibit them as valid monotones in a resource theory of quantum correlations with free operations restricted to the appropriate local channels.
\end{abstract}


\maketitle                           
\section{Introduction\label{sec:intro}}

For a long time, entanglement was believed to be the main resource for the practical implementation of quantum information processing. However, at present, it is widely recognized that entanglement is not the unique asset which can be exploited in quantum protocols in order to outperform their classical counterpart. Several results indicate that in some quantum tasks the processing improvements are due to correlations of a quantum nature different from entanglement \cite{Knill98}-\cite{Lanyon08}. \par

The prominent role of such (general) quantum correlations (QC) in the efficient realization of a number of tasks, has led to the introduction of several and different measures of QC, yet at present there is no general consensus regarding which is the most suitable and practical measure to be used in an arbitrary composite quantum system. Due to the subtle nature of quantumness, no single quantity captures all its essential features, and extensive effort has been dedicated in order to characterize it in varied ways.

One of the most widely used measure of quantum correlations in bipartite systems is the 
so-called (standard) quantum discord (QD)~\cite{OZ02,HV01}, which essentially quantifies the discrepancy between the quantum versions of two classically equivalent expressions for the mutual information. Although from a conceptual point of view QD is of relevance in assessing possible non-classical resources for information processing, it exhibits some practical drawbacks. For example, at this moment there is no straightforward criterion to verify the presence of nonzero discord in a given arbitrary bipartite quantum state. Besides, as the evaluation of QD involves an optimization procedure, analytical results are known only in some particular cases~\cite{Lang10}-\cite{Luo08b}. Furthermore, in general, calculation of quantum discord is a hard task, since the optimization procedure needs to be done by means of a sweep over a complete set of measurements performed over one of the subsystems \cite{Huang14}. \par

With the aim of finding alternative ways to quantify quantum correlations, several measures other than QD have been proposed \cite{HuFan15}-\cite{Guo16}, and the set of desirable properties that \textit{bonafide} measures of correlations should satisfy have been thoroughly discussed. Of particular importance for this work are the noncommutativity-based measures of QC introduced in \cite{Guo16} and \cite{MBOLVH17}, and some criteria presented in \cite{Brodutch12} and \cite{ABC16}. In particular, we focus on those criteria that are of relevance in the context of resource theories \cite{ABC16}, in which any well-behaved measure of a \textit{resource}, in this case QC, should be monotonically non-increasing under the action of certain  (so-called \textit{free}) operations. For a bipartite ($A$+$B$) system, and considering QC with respect to subsystem $A$ only, these operations are the Local Commutative Preserving Operations (LCPO) on $A$,\footnote{A LCPO corresponds to a map $\Delta[\cdot]$ that is completely positive trace preserving, and preserves the commutativity \cite{ABC16}, that is,
	$[\Delta[\rho],\Delta[\sigma]]=0 \ \ \forall\;\rho,\sigma \; \textrm{such that} \; [\rho,\sigma]=0.$
}  and local operations on $B$. Because of this, in this paper we focus on demonstrating the monotonously non-increasing behaviour of the non-commutativity based measure under these particular local operations.  

The paper is organized as follows. In Sec. \ref{sec:theory} we outline the basic theoretical background for our analysis, including the definition of asymmetrical (one-sided) quantum correlations, the properties that suitable QC measures should satisfy according to \cite{ABC16}, and the definition of the aforementioned measures of QC based on non-commutativity. In Sec. \ref{sec:main} we present our main results, demonstrating the non-increasing behaviour of such measures under any LCPO on subsystem $A$ for \textit{arbitrary} bipartite states, and under any local operation on subsystem $B$ for the \textit{restricted} family of Bell-diagonal states. Finally, some conclusions are drawn in Sec. \ref{sec:conclusions}.

\section{Theoretical framework\label{sec:theory}}

\subsection{One-sided quantum correlations in bipartite systems}\label{sec:Jvedral}

Let us consider a bipartite system $A+B$ with finite-dimensional Hilbert space $\mathcal{H}_{AB}=\mathcal{H}_A\otimes\mathcal{H}_B$, in a quantum state $\rho$. If the system is prepared in a pure state $\rho = \operdef{\psi_{AB}}{\psi_{AB}}$, correlations between the subsystems arise only when $|\psi_{AB}\rangle$ is an entangled (non-factorizable) state. That is, entanglement is the only possible form of (quantum) correlations in pure bipartite states. In contrast, if the system is prepared in a mixed state $\rho$, separable (unentangled) states can be endowed with nonclassical correlations, thus providing a subtler and richer situation compared with the pure-state case.\par

The set of separable states $\mathcal{S}_{AB}$ of bipartite quantum systems is composed of density operators $\rho \in \mathcal{D}_{AB}$ (with $\mathcal{D}_{AB}$ the convex set of density operators acting on $\mathcal{H}_{AB}$), that decompose as a convex sum of product states $\rho_A \otimes \rho_B$, where $\rho_{A(B)}$ refers to the (reduced) quantum state of subsystem $A$ ($B$). In other words, the set $\mathcal{S}_{AB}$ is defined as  
\begin{equation} \label{separables}
\mathcal{S}_{AB} := \{ \rho\; | \ \rho=\sum_i \ p_i \ \rho_A^i \otimes \rho_B^i \},
\end{equation}
where $\{p_i\}$ is a probability distribution satisfying $\sum_i p_i =1$. Any state $\rho$ that does not admit a decomposition of the form given by Eq. \eqref{separables} is said to be non-separable, or equivalently, entangled. 

Even though states in $\mathcal{S}_{AB}$ do not exhibit entanglement, non-classical correlations may be present in them. In order to determine the form of those quantum correlated states, we first establish the structure of the classically correlated ones. Specifically, classical correlations arise in states in which both subsystems $A$ and $B$ can be effectively described (or simulated) by classical systems, whose accessible states are distinguishable. Therefore, whenever all $\rho_A^i$ and $\rho_B^i$ in Eq. (\ref{separables}) reduce to projectors in some orthonormal basis $\{\left| i _A\right>\in \mathcal{H}_A \}$, and $\{\left| j _B\right>\in \mathcal{H}_B\}$, respectively, any correlation present is of classical nature. The corresponding state $\rho$ is said to be classical-classical, and belongs to the set of classical-classical states, defined according to 
\begin{equation}
\mathcal{C}_{AB} := \{ \rho  \;|\; \rho=\sum_{i,j} \ p_{ij} \ \left| i_A \right> \left< i_A \right|  \otimes  \left| j_B \right> \left< j_B \right| \},
\end{equation}
with $\{p_{ij}\}$ a joint probability distribution. 

States not contained in $\mathcal{C}_{AB}$ exhibit non-classical correlations of some kind. In particular, all separable states in which subsystem $A$ is classical but $B$ is not, give rise to the set of ($A$)classical-($B$)quantum states:
\begin{equation}\label{C_A}
\mathcal{C}_{A} := \{ \rho  \; | \; \rho=\sum_{i} \ p_{i} \ \left| i_A \right> \left< i_A \right|  \otimes  \rho_B^i \}.
\end{equation}
The states contained in $\mathcal{C}_{A}$ are thus said to be classically correlated with respect to A, and quantumly correlated with respect to B. In a similar way, exchanging $A\leftrightarrow B$ in the above expressions we can also define the set  $\mathcal{C}_B$ of ($A$)quantum-($B$)classical states.\par

In order to quantify the amount of asymmetric, one-sided, non-classical correlations in a given state $\rho$, several measures have been constructed. Although a unique set of criteria that appropriate positive and real functions $Q_A$ --measuring the quantum correlations with respect to subsystem $A$-- should satisfy is still an open problem, as discussed in \cite{Brodutch12} and \cite{ABC16}, reasonable \textit{necessary} properties are: 
\begin{enumerate}[label=\roman*)]
	\item $Q_A(\rho)=0$ if $\rho \in \mathcal{C}_A$; \label{cond 1}
	\item $Q_A(\rho)=Q_A(U_A\otimes U_B\,\rho\, U^{\dagger}_A\otimes U^{\dagger}_B)$ for all unitary operations $U_A, U_B$. (Invariance under local unitary operations); \label{cond 2}
	\item $Q_A(\ket{\psi_{AB}})=\mathcal{E}(\ket{\psi_{AB}})$, with $\mathcal{E}$ an appropriate entanglement measure. (QC reduces to an entanglement measure for pure states). \label{cond 3}
\end{enumerate}
Additional requirements for $A$-sided QC measures are \cite{ABC16}: 
\begin{enumerate}[label=\roman*),resume]
	\item $Q_A(\Delta_A \otimes \mathbb{I}_B[\rho]) \leq Q_A(\rho)$, where $\Delta_A$ is any LCPO on subsystem $A$. ($A$-sided quantum correlations are monotonously non-increasing under LCPO on $A$); \label{CA3}
	\item $Q_A(\mathbb{I}_A\otimes \Lambda_B\,[\rho]) \leq Q_A(\rho)$, where $\Lambda_B$ is any local operation on subsystem $B$. ($A$-sided quantum correlations are monotonously non-increasing under local operations on subsystem $B$).\label{CA2}
\end{enumerate}

\subsection{Non-commutativity measures of quantum correlations}

For the bipartite system under consideration, the state $\rho$ can always be expressed as
\beq\label{rho}
\rho=\sum_{i,j}  A_{ij}\otimes |i_B\rangle\langle j_B|,
\eeq
\noindent where $\{|i_B\rangle\}$ stands for an orthonormal basis of $\mathcal{H}_B$, and 
\beq \label{As}
A_{ij}=\mathrm{Tr}_B[(\mathbb{I}_A\otimes|j_B\rangle\langle i_B|)\rho]=\langle i_{B}|\rho|j_{B}\rangle.
\eeq 
With the operators $A_{ij}$ just defined, Guo \cite{Guo16} introduces the following measure of QC (with respect to subsystem $A$) : 
\beq
D_{A}(\rho):=\sum_{\Omega}||[A_{ij},A_{kl}]||_2,\label{Dtraza}
\eeq
where $||\cdot||_2$ is the Hilbert-Schmidt norm, $||A||_2=\sqrt{\mathrm{Tr}(A^{\dagger}A)}$, and $\Omega$ represents the set of all the possible pairs (regardless of the order). 

As discussed in \cite{Guo16}, $D_{A}(\rho)$ satisfies the requirements \ref{cond 1} and \ref{cond 2} above. Yet, as shown in \cite{MBOLVH17}, it fails in reducing to a satisfactory measure of entanglement for pure states (requirement \ref{cond 3}). In order to surmount this difficulty, the following quantity has been proposed as a more appropriate measure of $A$-sided quantum correlations \cite{MBOLVH17}: 
\beq\label{dnos}
d_{A}(\rho):=\min_{\mathcal{R}}D_{A}(\rho),
\eeq
where the minimum is taken over all representations of the state $\rho$. This improved measure of QC based on non-commutativity, thus satisfies requirements \ref{cond 1}-\ref{cond 3}. In what follows, we show that it always complies with condition \ref{CA3}, and that it satisfies condition \ref{CA2} whenever an important (yet restricted) family of two-qubit states is considered.

\section{Verification of conditions \ref{CA3} and \ref{CA2}}\label{sec:main}

\subsection{Criterion \ref{CA3}: Non-increasing behaviour under the action of any LCPO map on $A$}
Let us consider a map $\Delta_A$ acting on $A$, and the transformed state $\rho'=\left(\Delta_A\otimes\mathbb{I}_{B}\right)[\rho]$. Resorting to Eq. (\ref{As}) we have
\begin{align}\label{15}
A'_{ij}=\textrm{Tr}_B\left[\mathbb{I}_A\otimes \left|j_B\right>\left<i_B\right| \rho' \right]=\bra{i_B}\rho'\ket{j_B},
\end{align}
where $\rho'=\sum_{ij} \Delta_A\left[A_{ij}\right] \otimes \left|i_B\right> \left<j_B\right|.$
Thus, from Eq. (\ref{rho}) we can identify $A'_{ij}=\Delta_A\left[A_{ij}\right]$,
and consequently the measure $D_A(\rho')$ involves the commutators 
\begin{align}\label{19}
\left[ A'_{ij},A'_{kl} \right]=\left[ \Delta_A[A_{ij}],\Delta_A[A_{kl}] \right].
\end{align}

We now consider that $\Delta_A$ is a LCPO map. If $\dim \mathcal{H}_A=2$, then every LCPO map on $A$ is either a \textit{unital} or a \textit{completely decohering} map \cite{ABC16ref65}, which in turn is also unital.\footnote{Recall that a map $\Phi[\cdot]$ is said to be unital if $\Phi[\mathbb{I}]=\mathbb{I}$, whereas a completely decohering map $\Phi[\cdot]$ is such that $\Phi[\rho]=\sum_ip_i \ket{i}\bra{i}$, for some orthonormal basis $\{\ket{i}\}$ and (state-dependent) probabilities $\{p_i\}$.} Therefore, for $\dim \mathcal{H}_A=2$, every $\Delta_A$ is a unital map. If instead $\dim \mathcal{H}_A>2$, every LCPO map is an \textit{isotropic} or a completely decohering map \cite{ABC16ref63,ABC16ref66}. For these reasons in what follows we analyze the commutators (\ref{19}) considering separately the cases $\dim \mathcal{H}_A=2$ and $\dim \mathcal{H}_A>2$.

\subsubsection{The case $\dim \mathcal{H}_A=2$. Unital maps}
We start by expressing the $2\times 2$ matrices $\{A_{ij}\}$ and $\{A'_{ij}\}$ in terms of the $2\times 2$ identity matrix  $\mathbb{I}_2$, and $\vect \sigma=(\sigma_1,\sigma_2,\sigma_3)$, with $\{\sigma_i\}$ the Pauli matrices, thus getting
\begin{align}
&A_{ij}=\vect {d}_{ij}\cdot(\mathbb{I}_2,\vect \sigma), \\
&A'_{ij}=\vect {e}_{ij}\cdot(\mathbb{I}_2,\vect \sigma),\label{aprima}
\end{align}
where $\vect {d}_{ij}=(d^0_{ij}, {d}^1_{ij},{d}^2_{ij},{d}^3_{ij})$ and $\vect {e}_{ij}=(e^0_{ij}, {e}^1_{ij},{e}^2_{ij},{e}^3_{ij})$ are in general complex vectors. 

The commutator $[A_{ij},A_{kl}]$ thus reads
\beq
[A_{ij},A_{kl}]=\beta_{ijkl}^{(12)}\sigma_3 +\beta_{ijkl}^{(31)}\sigma_2 + \beta_{ijkl}^{(23)}\sigma_1,
\eeq
where we defined $\beta_{ijkl}^{(mn)}=2i\left(d_{ij}^md_{kl}^n-d_{ij}^nd_{kl}^m\right)$. This gives finally 
\begin{align}\label{AAs}
\begin{Vmatrix}
[A_{ij},A_{kl}]
\end{Vmatrix}^2_2=2\left(\left|\beta_{ijkl}^{(12)}\right|^2+ \left|\beta_{ijkl}^{(31)}\right|^2 + \left|\beta_{ijkl}^{(23)}\right|^2 \right).
\end{align}
Resorting to Eq. (\ref{aprima}), equivalent expressions are obtained for the matrices $\{A'_{ij}\}$:
\begin{align}\label{A's}	
\begin{Vmatrix}
[A'_{ij},A'_{kl}]
\end{Vmatrix}^2_2=2\left(\left|\eta_{ijkl}^{(12)}\right|^2+ \left|\eta_{ijkl}^{(31)}\right|^2 + \left|\eta_{ijkl}^{(23)}\right|^2 \right),
\end{align}
with $\eta_{ijkl}^{(mn)}=2i\left(e_{ij}^me_{kl}^n-e_{ij}^ne_{kl}^m\right)$. 

Now, any linear, trace and positive preserving map $\Phi$ in a bidimensional space that maps the operator $\hat M=\vect {m}\cdot(\mathbb{I}_2,\vect \sigma)$ into  $\hat M'=\vect {m}'\cdot(\mathbb{I}_2,\vect \sigma)$, is completely characterized by the following transformation on the corresponding vectors: 
\begin{equation}\label{transf}
\vect {m}'=\mathbb{T}\vect {m},
\end{equation}
where
\begin{align} \label{matrixT}
\mathbb{T}=\begin{bmatrix}
1&\vect {0} \\
{\vect {t}}^\top&T
\end{bmatrix},
\end{align}
with $\vect {0}=(0,0,0)$, $\vect {t}=(t_1,t_2,t_3)$, and $T$ the $3\times 3$ matrix $T=\textrm{diag}(\lambda_1,\lambda_2, \lambda_3)$ such that 
\beq \label{condL}
|\lambda_k|\leq 1-|t_k|\leq 1.
\eeq
This last condition results form the positivity-preserving condition \cite{RSW}. In addition, for unital maps it holds that $\vect {t}=\vect {0}$. 

In the present (unital) case, Eq. (\ref{transf}) becomes $\vect e_{ij}=\mathbb{T}\vect d_{ij}$, with $\vect {t}=\vect {0}$, whence $\eta_{ijkl}^{(mn)}=\lambda_m\lambda_n\beta_{ijkl}^{(mn)}$. Resorting to Eq. (\ref{condL}) gives thus $|\eta_{ijkl}^{(mn)}|^2\leq |\beta_{ijkl}^{(mn)}|^2$, and comparison of Eqs. (\ref{AAs}) and (\ref{A's}) results in
\beq\label{desA}
||[A'_{ij},A'_{kl}]||_2 \leq ||[A_{ij},A_{kl}]||_2.
\eeq
This inequality implies that under a LCPO (unital) operation on the qubit $A$, $D_{A}(\rho')\leq D_{A}(\rho)$, hence Guo's measure, and consequently the measure (\ref{dnos}), does not increase under LCPO whenever $\dim \mathcal{H}_A=2$. 
\subsubsection{The case $\dim \mathcal{H}_A>2$. Completely decoherent and isotropic maps}

We now consider  $\textrm{dim}\mathcal{H}_A>2$, in which case any LCPO map is either a completely decohering map, or an isotropic one. The former case is trivial, since a completely decohering map on $A$ takes an arbitrary state $\rho$ into an $A$-classical state, element of $\mathcal{C}_{A}$. Since  $D_A(\rho)=d_A(\rho)=0$ whenever $\rho \in \mathcal{C}_{A}$, it follows straightforward that neither $D_A$ nor $d_A$ increase under a completely decoherent map on subsystem $A$. Therefore, in what follows we focus on isotropic maps acting on $A$.\par
An isotropic map $\Phi^{\textrm{iso}}$ can be represented as
\begin{align}\label{iso}
\Phi^{\textrm{iso}}[\rho]=p\Gamma[\rho] + (1-p)\frac{\mathbb{I}_d}{d}\textrm{Tr}\,\rho,
\end{align}
with $d=\dim \mathcal{H}$, and $\Gamma$ a unitary or antiunitary operation.\cite{ABC16,ABC16ref63,ABC16ref66} As explained in detail in \cite{Broetal17}, for the map $\Phi^{\textrm{iso}}$ to be completely positive, the parameter $p$ must be such that $\frac{-1}{d^{2}-1}\leq p \leq 1$ whenever $\Gamma$ is unitary, and $\frac{-1}{d-1}\leq p \leq \frac{1}{d+1}$ in the case $\Gamma$ is antiunitary. Notice that in both cases $p^2\leq1$.\par
From Eq. (\ref{19}) with $\Delta_A=\Phi^{\textrm{iso}}$ we get
\begin{align}\label{21}
[ A'_{ij},A'_{kl} ]=p^2[\Gamma[A_{ij}],\Gamma[A_{kl}]].
\end{align}
For the unitary case we write $\Gamma[\rho]=U\rho U^\dagger$, then
\begin{align}
[ A'_{ij},A'_{kl}]=p^2 U [ A_{ij},A_{kl} ] U^\dagger,
\end{align}
and therefore
\beq
||[A'_{ij},A'_{kl}]||_2 =p^2 ||[A_{ij},A_{kl}]||_2\leq||[A_{ij},A_{kl}]||_2.
\eeq
As stated below Eq. (\ref{desA}), this implies that 
\beq\label{noincrDd}
D_A(\rho')\leq D_A(\rho),\quad d_A(\rho')\leq d_A(\rho).
\eeq

For the antiunitary case we put $\Gamma[\rho]=U\rho^\top U^\dagger$, thus
\begin{align}
[ A'_{ij},A'_{kl} ]=p^2 U [A_{ij}^\top,A_{kl}^\top] U^\dagger, 
\end{align}
so that resorting to the invariance of the Hilbert Schmidt norm under the transposition of its argument we arrive at
\beq
||[A'_{ij},A'_{kl}]||_2 =p^2 ||[A_{ij}^\top,A_{kl}^\top]||_2\leq||[A_{ij},A_{kl}]||_2,
\eeq
and we are finally led to the non-increasing behaviour of $D_A$ and $d_A$ (Eq. (\ref{noincrDd})) under isotropic maps on $A$.

\subsection{Criterion \ref{CA2}: Non-increasing behaviour under the action of any quantum local operation on $B$ }

The criterion will now be proved considering $\dim\mathcal{H}_A=\dim\mathcal{H}_B=2$, and restricting ourselves to states $\rho$ such that $\rho_A$ and $\rho_B$ are maximally mixed. That is, we analyze criterion \ref{CA2} for Bell-diagonal states:
\begin{align}\label{rhom}
\rho=\frac{1}{4}\left(\mathbb{I}_A\otimes\mathbb{I}_B+\sum_k c_k \sigma_k \otimes \sigma_k\right),
\end{align}
with real constants $c_k$.
Equation (\ref{As}) gives in this case:
\begin{align}
A_{ij}=\frac{1}{4}\mathbb{I}_A\delta_{ij}+\frac{1}{4}\sum_k c_k \sigma_k^{ij} \sigma_k,
\end{align}
where $\sigma_k^{ij}=\bra{i_B}\sigma_k \ket{j_B}$. The commutators thus result
\begin{align}
&\left[A_{ij},A_{kl}\right]=\frac{i}{8}\Big[ c_1c_2\alpha^{(12)}_{ijkl}\sigma_3+c_1c_3\alpha^{(31)}_{ijkl}\sigma_2+c_2c_3\alpha^{(23)}_{ijkl}\sigma_1  \Big],
\end{align}
with $\alpha_{ijkl}^{(mn)}=\sigma_m^{ij} \sigma_n^{kl} -  \sigma_n^{ij} \sigma_m^{kl}$. This gives
\begin{align}\label{26}
&\begin{Vmatrix}
\left[A_{ij},A_{kl}\right]
\end{Vmatrix}_2^2=\frac{1}{2^5}\Big( 
\left|c_1c_2\right|^2\left|\alpha_{ijkl}^{(12)} \right|^2 +\left|c_1c_3\right|^2\left|\alpha_{ijkl}^{(31)} \right|^2 + 
\left|c_2c_3\right|^2\left|\alpha_{ijkl}^{(23)} \right|^2\Big).
\end{align}
We now consider the transformed state $\rho'=(\mathbb{I}_A\otimes\Phi)\rho$, obtained after applying a completely positive trace preserving map $\Phi$ on subsystem $B$, 
\begin{align}\label{rhom2}
\rho'=\frac{1}{4}\left(\mathbb{I}_A\otimes\Phi[\mathbb{I}_B]+\sum_k c_k \sigma_k \otimes \Phi[\sigma_k]\right).
\end{align}
Denoting with $\vect{x}_j$ the vector that has $1$ in the $j$-th position ($j=1,2,3$), and zeros in the two remaining entries, we can define $\vect{s}_j=(0,\vect{x}_j)$ and write $\vect{s}_j\cdot(\mathbb{I},\vect{\sigma})=(0,\vect{x}_j)\cdot(\mathbb{I},\vect{\sigma})=\sigma_j$. With this, and writing $\mathbb{I}_B=\vect{d} \cdot(\mathbb{I},\vect{\sigma})$ with $\vect{d}=(1,\vect{0})$, we resort to Eq. (\ref{transf}) and obtain
\begin{align}
&\Phi[\mathbb{I}_B]=\mathbb{I}'_B=(\mathbb{T}\vect d)\cdot(\mathbb{I},\vect{\sigma})=(1,\vect{t}) \cdot(\mathbb{I},\vect{\sigma}), \label{45} \\
&\Phi[\sigma_j]=\sigma'_j=(\mathbb{T}\vect{s}_j)\cdot(\mathbb{I},\vect{\sigma})=\lambda_j[\vect{s}_j\cdot(\mathbb{I},\vect{\sigma})]=
\lambda_j \sigma_j. \label{46}
\end{align}
Using Eqs.  (\ref{rhom2})-(\ref{46}) we compute the commutators resulting after the action of the map,
\begin{align}\label{26b}
&\begin{Vmatrix}
\left[A'_{ij},A'_{kl}\right]
\end{Vmatrix}_2^2=\frac{1}{2^5}\Big( 
\left|\lambda_1\lambda_2\right|^2\left|c_1c_2\right|^2\left|\alpha_{ijkl}^{12} \right|^2 +\nonumber \\
&+\left|\lambda_1\lambda_3\right|^2\left|c_1c_3\right|^2\left|\alpha_{ijkl}^{31} \right|^2 + 
\left|\lambda_2\lambda_3\right|^2\left|c_2c_3\right|^2\left|\alpha_{ijkl}^{23} \right|^2\Big)
\end{align}
Finally, given that $|\lambda_j|\leq1$ (Eq. (\ref{condL})), comparison of Eqs. (\ref{26}) and (\ref{26b}) leads us to 
\begin{align}
\begin{Vmatrix}
\left[A'_{ij},A'_{kl}\right]
\end{Vmatrix}_2^2 \leq \begin{Vmatrix}
\left[A_{ij},A_{kl}\right]
\end{Vmatrix}_2^2,
\end{align}
hence to Eq. (\ref{desA}), which demonstrates the non-increasing behaviour of $D_A$ and $d_A$ under local operations on $B$ for Bell-diagonal states.

\section{Concluding remarks\label{sec:conclusions}}
In this work we studied the behaviour of the non-commutativity measures of QC (\ref{Dtraza}) and (\ref{dnos}) under the action of local operations either on subsystem $A$ or $B$. Specifically, we demonstrated that irrespective of the particular bipartite system, $D_A$ and $d_A$ are non-increasing under any Local Commutative Preserving Operation on subsystem $A$, a feature that strengthens the benefits of resorting to such measures. In this sense, though we have only analyzed non-commutativity measures defined in terms of the Hilbert-Schmidt norm, it seems plausible to carry out a similar analysis involving any arbitrary norm satisfying some basic properties (as for instance, the invariance under the transposition of its argument). We also showed that when considering (two-qubit) Bell-diagonal states, $D_A$ and $d_A$ do not increase under arbitrary local operations on subsystem $B$. Though this last property has only been proved for a restricted set of states (which at this state may limit the measures as a faithful correlation monotone in a resource theory), and its extension to arbitrary bipartite states still remains pending, our findings represent an important advance regarding the potentialities of the measures, and show that they shape up as a promising option for a quantum correlation resource theory.

\begin{acknowledgements}
	D. B., and A. P. M. acknowledge the Argentinian agency SeCyT-UNC and CONICET for financial support. D. B. has a fellowship from CONICET. A. V. H.  gratefully acknowledges financial support from DGAPA, UNAM through project PAPIIT IA101918.
\end{acknowledgements}


\begin{thebibliography}{}
	
	\bibitem{Knill98} Knill, E., Laflamme, R.: Power of One Bit of Quantum Information. Phys. Rev. Lett. {\bf 81}, 5672 (1998)
	\bibitem{LCNV} Laflamme, R., Cory, D. G., Negrevergne, C. \& Viola, L.: NMR quantum information processing and entanglement. Quantum. Inf. and Comp. {\bf 2}, 166 (2002)
	\bibitem{Braun99} Braunstein, S. L., Caves, C. M., Jozsa, R., Linden, N., Popescu, S., Schack R.: Separability of very noisy mixed states and implications for NMR Quantum computing. Phys. Rev. Lett. {\bf 83}, 1054 (1999)
	\bibitem{Meyer00} Meyer, D., A.: Sophisticated Quantum Search Without Entanglement. Phys. Rev. Lett. {\bf 85}, 2014 (2000)
	\bibitem{Datta05} Datta, A., Flammia, S., T., Caves, C., M.: Entanglement and the power of one qubit. Phys. Rev. A {\bf 72}, 042316 (2005)
	\bibitem{Datta07} Datta A., Vidal, G.: Role of entanglement and correlations in mixed-state quantum computation. Phys. Rev. A {\bf 75}, 042310 (2007)
	\bibitem{Datta08} Datta, A., Shaji, A., Caves, C. M.: Quantum Discord and the Power of One Qubit. Phys. Rev. Lett. {\bf 100}, 050502 (2008)
	\bibitem{Lanyon08} Lanyon, B., P., Barbieri, M., Almeida M., P., White, A., G.: Experimental Quantum Computing without Entanglement. Phys. Rev. Lett. {\bf 101}, 200501 (2008)
	\bibitem{OZ02} Ollivier, H., Zurek,  W., H.: Quantum Discord: A Measure of the Quantumness of Correlations. Phys. Rev. Lett {\bf 88}, 017901 (2001)
	\bibitem{HV01} Henderson, L., Vedral, V.: Classical, quantum and total correlations. J. Phys. A {\bf 34}, 6899 (2001)
	\bibitem{Lang10} Lang, M., D., Caves, C., M.: Quantum Discord and the Geometry of Bell-Diagonal States. Phys. Rev. Lett. {\bf 105}, 150501 (2010)
	\bibitem{Cen11} Cen, L.-X., Li, X., Q., Shao, J., Yan, Y., J.: Quantifying quantum discord and entanglement of formation via unified purifications. Phys. Rev. A {\bf 83}, 054101 (2011)
	\bibitem{Adesso10} Adesso, G., Datta, A.: Quantum versus Classical Correlations in Gaussian States. Phys. Rev. Lett. {\bf 105}, 030501 (2010); Giorda, P. and Paris, M., G., A.: Gaussian Quantum Discord. ibid. 105, 020503 (2010)
	\bibitem{Ali10} Ali, M., Rau, A., R., P., Alber, G.: Quantum discord for two-qubit X states. Phys. Rev. A {\bf 81}, 042105 (2010); see also Ali M. , Rau, A., R., P., and Alber, G., ibid. 82, 069902(E) (2010)
	\bibitem{Shi11} Shi, M., Yang, W., Jiang, F., Du, J.: Quantum discord of two-qubit rank-2 states. J. Phys. A: Math. Theor. \textbf{44}, 415304 (2011)
	\bibitem{Chen11} Chen, Q., Zhang, C., Yu, S., Yi, X., X., Oh, C., H.: Quantum discord of two-qubit X states. Phys. Rev. A \textbf{84}, 042313 (2011)
	\bibitem{Lu11} Lu, X., M., Ma, J., Xi, Z., Wang, X.: Optimal measurements to access classical correlations of two-qubit states. Phys. Rev. A {\bf 83}, 012327 (2011)
	\bibitem{Giro12} Girolami, D., Adesso, G.: Quantum discord for general two-qubit states: Analytical progress. Phys. Rev. A {\bf 83}, 052108 (2011)
	\bibitem{Li11} Li, B., Wang, Z., X., Fei, S., M.: Quantum discord and geometry for a class of two-qubit states. Phys. Rev. A \textbf{83}, 022321 (2011)
	\bibitem{Huang13} Huang, Y.: Quantum discord for two-qubit X states: Analytical formula with very small worst-case error. Phys. Rev. A {\bf 88}, 014302 (2013)
	\bibitem{Luo08b} Luo S.: Quantum discord for two-qubit systems. Phys. Rev. A {\bf 77}, 042303 (2008)
	\bibitem{Huang14} Huang, Y.: Computing quantum discord is NP-complete. New J. Phys., \textbf{16}, 033027 (2014)
	
	\bibitem{HuFan15}Hu, M-L., Fan, H.: Measurement-induced nonlocality based on the trace norm. New J. Phys. {\bf 17}, 033004 (2015)
	\bibitem{HuFan12}Hu, M-L., Fan, H.: Dynamics of entropic measurement-induced nonlocality in structured reservoirs. Ann. Phys. {\bf 327}, 2343 (2012)
	\bibitem{DVB10} Daki\'c B., Vedral V., Brukner C.: Necessary and sufficient condition for nonzero quantum discord. Phys. Rev. Lett. {\bf 105}, 190502 (2010)
	\bibitem{Brod10} Brodutch A., Terno D. R.: Quantum discord, local operations, and Maxwell's demons. Phys. Rev. A \textbf{81}, 062103 (2010)
	\bibitem{Paula13} Paula F. M., de Oliveira T. R., Sarandy M. S.: Geometric quantum discord through the Schatten 1--norm. Phys. Rev. A \textbf{87}, 064101 (2013)
	\bibitem{Spehner14a} Spehner D., Orszag M.: Geometric quantum discord with Bures distance. New J. Phys. \textbf{15}, 103001 (2013)
	\bibitem{Spehner14b} Spehner D., Orszag M.: Geometric quantum discord with Bures distance: the qubit case. J. Phys. A: Math. Theor.\textbf{47}, 035302 (2014)
	\bibitem{Jakob14} Jak\'{o}bczyk L.: Spontaneous emission and quantum discord: Comparison of Hilbert--Schmidt and trace distance discord. Phys. Lett. A \textbf{378}, 3248-3253 (2014)
	\bibitem{KheAM16}Kheirollahi, A., Akhtarshenas, S. J., Mohammadi. H.: Quantifying nonclassicality of correlations based on the concept of nondisruptive local state identification. Quantum Inf. Process. {\bf 15}, 1585 (2016)
	\bibitem{LuoGD10} Luo S., Fu S.: Hybrid potential model of the $\alpha$-cluster structure of 212Po. Phys. Rev. A \textbf{82}, 034302 (2010)
	\bibitem{Guo16} Guo Y.: Non-commutativity measure of quantum discord. Sci. Rep. {\bf 6}, 25241 (2016)
	\bibitem{MBOLVH17} Majtey, A., P., Bussandri, D., G., Ossan T., G., Lamberti P., W., Valdés-Hernández A.: Problem of quantifying quantum correlations with non-commutative discord. Quantum Inf. Process {\bf 16}, 226 (2017)
	\bibitem{Brodutch12} Brodutch, A., Modi, K.: Criteria for measures of quantum correlations. Quantum Inf. Comput. 12, 721?742 (2012)
	\bibitem{ABC16} Adesso, G., Bromley, T., R., Cianciaruso, M.: Measures and applications of quantum correlations. J. Phys. A: Math. Theor. {\bf 49}, 473001 (2016)
	\bibitem{ABC16ref65} Streltsov, A., Kampermann, H., Bru\ss, D.: Behavior of Quantum Correlations under Local Noise. Phys. Rev. Lett. {\bf 107}, 170502 (2011)
	\bibitem{ABC16ref63} Hu, X., Fan, H., Zhou, D., L., Liu, W., M.: Necessary and sufficient conditions for local creation of quantum correlation. Phys. Rev. A. {\bf 85}, 032102 (2012)
	\bibitem{ABC16ref66} Guo, Y., Hou, J.: Necessary and sufficient conditions for the local creation of quantum discord. J. Phys. A: Math. Theor. {\bf 46}, 155301 (2013)
	\bibitem{RSW} Ruskai, M., B., Szarek, S., Werner, E.: An analysis of completely-positive trace-preserving maps on M2. Linear Algebr. Appl. {\bf 347}, 159 (2002)
	\bibitem{Broetal17} Bromley, T. R., Silva, I. A., Oncebay-Segura, C. O., Soares-Pinto, D. O., R. deAzevedo, E., Tufarelli, T., Adesso, G.: There is more to quantum interferometry than entanglement. Phys. Rev. A {\bf 95}, 052313 (2017)
	
	
	
	
\end{thebibliography}

{}

\end{document}